# National, disciplinary and temporal variations in the extent to which articles with more authors have more impact: Evidence from a geometric field normalised citation indicator[1]

Mike Thelwall, Pardeep Sud
Statistical Cybermetrics Research Group, School of Mathematics and Computer Science, University of Wolverhampton, Wulfruna Street, Wolverhampton WV1 1LY, UK.
Tel. +44 1902 321470. Fax +44 1902 321478, Email: m.thelwall@wlv.ac.uk p.sud@wlv.ac.uk

The importance of collaboration in research is widely accepted, as is the fact that articles with more authors tend to be more cited. Nevertheless, although previous studies have investigated whether the apparent advantage of collaboration varies by country, discipline, and number of co-authors, this study introduces a more fine-grained method to identify differences: the geometric Mean Normalized Citation Score (gMNCS). Based on comparisons between disciplines, years and countries for two million journal articles, the average citation impact of articles increases with the number of authors, even when international collaboration is excluded. This apparent advantage of collaboration varies substantially by discipline and country and changes a little over time. Against the trend, however, in Russia solo articles have more impact. Across the four broad disciplines examined, collaboration had by far the strongest association with impact in the arts and humanities. Although international comparisons are limited by the availability of systematic data for author country affiliations, the new indicator is the most precise yet and can give statistical evidence rather than estimates.

## 1. Introduction

Cooperation in research is promoted by many funding agencies in the belief that collaborative research tends to have more impact. This originates from the theoretical argument that interdisciplinary collaboration is often necessary to solve important societal problems (Gibbons, Limoges, Nowotny, Schwartzman, Scott, & Trow, 1994) and is supported by studies showing that collaborative research is often more highly cited than comparable solo studies (e.g., Thurman & Birkinshaw, 2006). International collaboration seems to be also promoted for partly political purposes, such as the European Union funding programmes that require at least three different member states to be represented within a funding bid (EC, 2014). These initiatives have presumably contributed to an increase in research collaboration in most fields (Wuchty, Jones, & Uzzi, 2007). Nevertheless, the value of collaboration seems to vary between fields, nations and type (e.g., national vs. international) and so it is important to understand where it is beneficial so that it can be promoted when it is most useful and perhaps even discouraged when it is problematic.

Although, as discussed below, previous studies have assessed factors that influence the success of collaborations, at least as reflected in the citation counts of the resulting publications, it is difficult to get a clear understanding of differences between collaboration types. This is because citation counts are highly skewed (e.g., Seglen, 1992) and so comparing arithmetic mean citation rates between different types of articles is unreliable, needing large sample sizes to give reasonable statistical power. Moreover, analyses need to harness large sets of articles in order to reliably distinguish between the average impacts of sets of articles with similar properties (e.g., articles with two authors compared to articles with three authors). Hence a more precise method is needed to compare the effects of collaboration on collections of articles, such as the geometric mean. This may not be enough, however, since the geometric mean can only be applied to articles from a single year and field because of differences in average citation counts. Hence, an indicator is needed that combines the ability of the geometric mean to deal with skewed data with the ability of the Mean Normalized Citation Score (MNCS) (discussed below) to combine citation counts from multiple fields and years. In response, this article introduces a variant of the MNCS, the geometric MNCS (gMNCS), and applies it to assess the effect of field, year and country on the extent to which the average citation impact of collaborative research articles varies with the number of authors. Geometric variants of several standard bibliometric indicators have previously been proposed, following their initial introduction (Zitt, 2012). These include geometric journal impact factors (Thelwall & Fairclough, 2015) and a basic average citation indicator for individual subjects and years (Fairclough, & Thelwall, 2015).

## 2. Background

Academic collaboration is the combining together of the expertise of multiple people in the production of research (Katz & Martin, 1997). In practice, even apparently solo research projects are sometimes collaborative to some extent through informal discussions with colleagues and help from support staff. Whilst these are important parts of the research process, collaborations that combine a substantial amount of academic expertise from the contributors are of particular interest because of its promotion by funding agencies in the belief that it tends to produce better research. In practice, quantitative studies of collaboration almost always focus on work that leads to published findings (there are many qualitative exceptions, e.g.: Latour & Woolgar, 1979) and use the authorship list as a proxy for the set of people that have substantially contributed to a study. Other contributors are sometimes recognised in an acknowledgement (Cronin, McKenzie, & Stiffler, 1992; Cronin, 2001a) but these are rarely analysed on a large scale.

The authorship list is a simplification of the concept of collaboration because it may omit important contributors (ghost authorship: Gotzsche, Hróbjartsson, Johansen, Haahr, Altman, & Chan, 2007) and include non-contributors (gift/honorary authorship: Cronin, 2001b; Drenth, 1998; Smith, 1994). Scientists also do not have a uniform understanding of concept of research collaboration and frequently do not grant co-authorships to people that have helped in research (Laudel, 2002). Moreover, although the authors are normally assumed to have contributed equally, in most fields the first author probably contributes more than the others (Vinkler, 1993). This is not true in all fields, with exceptions including mathematics, business and economics (Levitt & Thelwall, 2013) and there is no agreed formula to estimate the likely relative contributions of authors based on their order in the authorship list. It is becoming more possible to detect the value of the different authors for a paper because some journals require specific information about individual contributions (Bates, Anić, Marušić, & Marušić, 2004) but this falls short of giving a formula to estimate the relative importance of each one. In this context, it seems reasonable to accept the simplification that all authors' contributions are equally important.

Academic collaboration leading to co-authorship can be of very different types. A common type is probably junior-senior co-authorship where the main author is a PhD student and the second author is their main supervisor. Here, the student may have done most of the work but the supervisor may have provided expertise in the form of ideas and overall guidance on topic areas and specific advice about the research design, methods, analysis, write-up and publication venue. The exact nature of the relationship may vary by discipline, however (e.g., Barnes & Randall, 2012). In contrast, some collaborations involve sets of experienced researchers that provide complementary expertise from different fields, subfields, or tasks (e.g., statistics, interviews), in order to conduct studies that they could not perform as well individually. Other collaborations may also be between researchers with essentially identical skill sets but with their combined insights helping to solve a problem that they could, in theory, have addressed individually. For a large scale bibliometric analysis of publications no method has yet been developed to distinguish between these types of contributions and so there is no alternative to treating all types of collaboration as the same. When interpreting the results, however, the different types of collaboration should be considered as possible explanations for any patterns found.

Many, but not all (Bornmann, Schier, Marx, & Daniel, 2012; Haslam, Ban, Kaufmann, Loughnan, Peters, Whelan, & Wilson, 2008), studies investigating the connection between collaboration and citation have found that articles with more authors tend to be more cited (e.g., Thurman & Birkinshaw, 2006; Vieira, & Gomes, 2010). Most articles cannot be easily generalised, however, due to a focus on a set of publications with a specific attribute, such as originating from a single university, country, journal or field. There have also been variations in the types of collaboration examined, from a course grained comparison of solo with collaborative research, to comparisons of types of collaboration (e.g., intra-institutional, international) and different numbers of authors.

Not all types of collaboration have equal apparent impact. It seems that collaboration is particularly likely to generate higher (arithmetic mean) impact research if the collaborators are from different countries (Didegah & Thelwall, 2013; Glänzel, 2001; Katz, & Hicks, 1997), except perhaps in the social sciences (Didegah & Thelwall, 2013), and for authors at prestigious universities (e.g., Gazni & Didegah, 2010). Domestic collaborations seem to have the same impact whether multiple institutions are involved or not, however (Didegah & Thelwall, 2013). Moreover, collaboration within an institution associates with lower impact papers that solo research in at least one field (Leimu & Koricheva, 2005). The advantage of collaboration can also vary by country. For example, a study of biochemical research found that collaboration with the USA associated with increased citation impact for authors from other countries, but collaboration with some other countries was associated with decreased citation impact (Sud & Thelwall, in press).

Focusing on a single country can reveal the extent to which collaboration is advantageous for that nation. For economics, collaborative research has been shown to be at least as highly cited as solo research across 18 different countries and 12 different states in the USA, although the extent to which this was true for 5 other different states depended on the indicator used (Levitt, & Thelwall, 2010). This example illustrates the importance of using the most precise indicator when comparing impact.

The impact of collaboration varies by field, as confirmed above by the finding that international collaboration is apparently not beneficial in the Social Sciences, in comparison to two other Web of Science (WoS) categories (Biology & Biochemistry and Chemistry, 2000-2009). In the humanities, solo research is particularly valued in the form of the monograph (Williams, Stevenson, Nicholas, Watkinson, & Rowlands, 2009), and collaboration does not generate more monograph citations (Thelwall & Sud, 2014). A study of 11,196 journal articles and reviews within WoS from 2000, 2003 and 2005 with at least one author from South Africa in 18 natural and life science fields found solo authored papers to be more cited than collaboratively authored papers in some cases (e.g., Psychiatry, Biochemistry) but the converse was true for others (Engineering, Plant Sciences) (Sooryamoorthy, 2009). No statistical tests were conducted, however, and no allowances were made for skewed citation count data, the citation counts from the different years were not normalised and the sample sizes were small and so the findings are indicative rather than robust. A large scale study of the articles, letters and notes (in the Web of Science predecessor) of 50 nations in 1995/96 compared eight different broad fields: Clinical Medicine; Biomedical Research; Biology; Chemistry; Physics, Mathematics; Engineering; Earth and Space Sciences (Glänzel, 2001). Citation counts were used with a three year citation window. An indicator was used to assess the extent to which internationally-co-authored papers attracted more citations than did domestically authored publications within each of the 8 fields (i.e., 50x8 calculations; see Table 3 on p. 89: Glänzel, 2001). Internationally collaborative articles tended to attract more citations than did domestically authored articles, whether collaborative or not, in most cases but the results did not distinguish between international and domestic collaboration (e.g., by excluding non-collaborative domestic articles) and the indicator used did not take into account skewed citation patterns. This article also found that collaborations between specific pairs of nations may be particularly advantageous for one or other of the collaborators in terms of attracting more citations than domestic research in the same field (Glänzel, 2001).

Regression analyses of citation data can help to check whether collaboration is likely to be a cause of articles being more cited or whether it is an indirect effect of other properties of an article that more directly associate with higher citation counts. For example, if collaboratively produced articles tended to have clearer abstracts and be cited more as a result of this, then a regression analysis that included both author counts and an abstract clarity metric could distinguish between the effects of both variables. A regression analysis of a large number of variables has confirmed that papers with more authors tended to be more cited, however, irrespective of the nature of the collaboration (e.g., national or international) (Didegah & Thelwall, 2013). This was found to be true for all three of the subject areas analysed (Biology & Biochemistry, Chemistry and Social Sciences).

Many of the studies reviewed above have explicitly or implicitly assumed that research that is more highly cited tends to be better, other factors being equal. The use of citation counts as a proxy for quality is problematic, however, and at best they are an indicator of academic impact. Moreover, citation comparisons may not be fair for collaborative research since the additional citations may accrue from the wider audience generated by the extra authors rather than from higher quality research (Wallace, Larivière, & Gingras, 2012). National biases within a citation index can also influence citation counts for authors based on their nationality (Van Leeuwen, Moed, Tijssen, Visser, & Van Raan, 2001) because of the tendency for national self-citations (Lancho-Barrantes, Guerrero-Bote, & de Moya-Anegón, 2013; see also: Thelwall & Maflahi, 2015) and publishing in the national literature. A second order effect can also come from collaborating countries that tend to cite each other (Lancho-Barrantes, Guerrero-Bote, & de Moya-Anegón, 2013). Some apparent advantages or disadvantages of collaboration may therefore be indirect effects of the coverage of the citation database used. Nevertheless, there is evidence from Italy that peer ratings of collaborative articles tend to be higher than peer ratings of solo articles (Franceschet, & Costantini, 2010), suggesting that collaboration tends to be advantageous overall.

# 3. The Geometric Mean Normalized Citation Score (gMNCS)

Field normalisation is important when analysing collections of academic articles because fields may have different average numbers of citations per article so that it is not reasonable to combine raw citation counts from different fields. The MNCS approach is to convert each citation count to a normalised version by dividing it by the mean number of citations per paper for the field and year of publication. Articles with a normalised score greater than 1 have therefore attracted more citations than average for their field and year. Hence, if the number of citations for an article in field $f$ and year $y$ is $c_{fy}$ then the normalised value is $c_{fy}/\overline{c_{fy}}$, where the mean is taken over all articles published in the field and year. After this normalisation, articles from different years and fields could reasonably be compared against each other or combined into sets for analyses of their properties (Waltman, van Eck, van Leeuwen, Visser, & van Raan, 2011a,b).

A problem with the MNCS is that citation counts are highly skewed (Thelwall, & Wilson, 2014a; Wallace, Larivière, & Gingras, 2009) and the arithmetic mean used in its formula is therefore an imprecise central tendency measure (Bland & Altman, 1996). Particularly, for smaller data sets, the arithmetic mean is vulnerable to having a non-trivial fraction of its value attributable to a single highly cited article. A solution to this problem is to use the geometric mean instead of the arithmetic mean, as previously noted in other contexts for citation data (Zitt, 2012). With a standard offset of 1 to deal with uncited articles, the geometric mean of a set of articles is $\exp\left(\frac{1}{n}\sum \log(1 + c_{fy})\right) - 1$, where the sum is over all articles from the field and year (Thelwall & Wilson, 2014b). Essentially, this transforms the data using a logarithmic transformation, calculates the arithmetic mean, and then uses the exponential function to undo the effect of the transformation. For convenience, a curve above a variable will be used here to denote the geometric mean $\check{c}$ and MNCS with the geometric mean instead of the arithmetic mean $c_{fy}/\widetilde{c_{fy}}$ will be called the geometric MNCS.

After normalisation as above, sets of articles could be reasonably compared. For example, if one set of articles was of solo-authored research and the other was of collaborative research then the average impact of the two sets could be compared to see which tended to attract the most citations for its field and year. Although this could be achieved, for example, by comparing the arithmetic mean of the set of solo articles with the arithmetic mean of the set of collaborative articles, this is not ideal because of the skewed nature of citation counts, even after normalisation. A more precise and more powerful way to compare the averages of the two sets would be to calculate the geometric mean of the normalised citation counts of each set. Here the formula would be $\exp\left(\frac{1}{n}\sum \log(1 + c_{fy}/\widetilde{c_{fy}})\right) - 1$, where $n$ is the number of articles in the set and the sum is over all the articles in the set (i.e., a double use of the geometric mean). Confidence intervals can be calculated using confidence interval formulae for the normal distribution by omitting the reverse transformation $\exp(\cdot) - 1$ stage, in which case the data should be approximately normally distributed (because citation counts fit the discretised lognormal distribution well: Evans, Hopkins, & Kaube, 2012; Radicchi, Fortunato, & Castellano, 2008; Thelwall & Wilson, 2014c), then applying the reverse transformation $\exp(\cdot) - 1$ to the confidence interval limits (Thelwall & Wilson, 2014a). This confidence interval is not symmetric about the central estimate and is an approximation since the transformed citation data is not exactly normal. This double geometric mean approach should make the average estimates as precise as possible, hence having the most power to distinguish between different sets of articles, as well as allowing their precision to be estimated so that the significance of any differences detected can be checked.

The problem of using the arithmetic mean in the MNCS has been previously identified and the median has been proposed as an alternative (Leydesdorff & Opthof, 2011). The geometric mean is more useful, however, because it is more fine grained and is not subject to sudden incremental increases with the addition of a single moderately cited article. In practice, however, the geometric MNCS suffers from other problems that are the same as those of the MNCS: the difficulty of effectively normalising multiply-classified articles; and reliance upon the classification scheme used to categorise the articles during normalisation (Leydesdorff & Opthof, 2011).

# 4. Research Questions

This article uses the geometric MNCS to assess the impact of collaboration on the average citation impact of research. The study is driven by the following research questions. Although they have previously been partially answered with different indicators, they are addressed again here for the new variant of average impact and with the hope of generating statistical evidence to answer each question.

- How does the average citation impact of collaboratively produced articles vary with the number of collaborators?
- Does the average citation impact of collaboratively produced articles vary with the country of the collaborators?
- Does the average citation impact of collaboratively produced articles vary over time?
- Does the average citation impact of collaboratively produced articles vary with discipline?

# 5. Data and Methods

The overall research design was to gather a large sample of articles from different disciplines and time periods in order to compare the average impact within each set based upon the number of authors and their nationalities. Scopus was chosen in preference to WoS because it seems to have greater coverage (Leydesdorff, de Moya-Anegón, & Nooy, in press), including wider international coverage (López-Illescas, de Moya-Anegón, & Moed, 2008) which is particularly important for international comparisons, although it has a bias towards English language publications (de Moya-Anegón, Chinchilla-Rodríguez, Vargas-Quesada, Corera-Álvarez, Muñoz-Fernández, González-Molina, & Herrero-Solana, 2007).

To investigate overall changes over time, the first data set included 25 varied Scopus categories from 2009 to 2015 in order to give wide coverage of scholarship: Animal Science and Zoology; Language and Linguistics; Biochemistry; Business and International Management; Catalysis; Electrochemistry; Computational Theory and Mathematics; Management Science and Operations Research; Computers in Earth Sciences; Finance; Fuel Technology; Automotive Engineering; Ecology; Immunology; Ceramics and Composites; Analysis; Anesthesiology and Pain Medicine; Biological Psychiatry; Assessment and Diagnosis; Pharmaceutical Science; Astronomy and Astrophysics; Clinical Psychology; Development; Food Animals; Complementary and Manual Therapy. This is a systematic sample of categories from Scopus with each being the third categories in a broad Scopus category (e.g., Animal Science and Zoology 1103 is the third category in the Agricultural and Biological Sciences 1100 broad category). The complete set is therefore reasonably representative of journals classified by Scopus with the exception that the (small) broad category Dentistry is not represented because its third category, Dental Hygene 3503, had no data in some years. The third category was a relatively arbitrary choice and the second or fourth could equally have been chosen. The first category is Miscellaneous in all cases and Veterinary has only four categories. The data was downloaded from Scopus between September 15 and 30, 2015. A fixed citation window was not used but all citations to date were included to give the maximum power for the data set. This is a limitation for comparisons over time, however, because differences between years could be caused by the variable citation windows or by underlying changes in research. The year 2015 was included despite it being incomplete and having very limited and therefore unreliable citation counts. This is because the policy makers that call for international comparisons typically need current data to assess the influence of recent policies and so, although not answering a research question, it is useful to include results from this year in order to confirm that older data is needed for reasonable results.

For each article, the country affiliations of the authors were extracted from Scopus, when present. Articles were categorised as purely national if only one country was recorded. This information was available for a minority of articles. The articles without nationality information were used to construct the world average for citations but were not included in any country set.

To investigate disciplinary differences, the second data set consisted of journal articles (excluding reviews and all other non-article types) in all subcategories of the Scopus sections: Arts and Humanities (13 subcategories); Business, Management and Accounting (10 subcategories); Chemistry (7 subcategories); and Pharmacology, Toxicology and Pharmaceutics (5 subcategories). These were selected relatively arbitrarily to represent four different types of scholarship: humanities, social science, medicine, and natural science. The data included up to 10000 articles for each year 2009-2013 and for each subcategory. The Scopus API returns a maximum of 5000 results per query and returns them in date order and so queries with up to 10000 matches can get complete results by submitting two versions of the query, with the second reversing the date sorting option. For years with over 10000 articles, the first 5000 and last 5000 in the year were retrieved using the Scopus sort by date option. Although it is not ideal to use a subset of articles, this subset is at least a balanced set of the most recent and oldest articles. The years 2009-2013 were chosen to give enough recent years for a large data set, but excluding the most recent two years for which there may be few citations. For each article, the Scopus-indexed citation counts were extracted between the $2^{nd}$ and $6^{th}$ of October 2015. The same information as for the first set was extracted.

National comparisons were made for nine countries, selected to include major research producers and other countries of interest: United States; Germany; France; Canada; Japan; China; United Kingdom; Italy; Russian Federation. Other than the major research nations represented, the inclusion of selected others is a relatively arbitrary choice and was made to follow an independent source: an international comparison made for the UK government (Elsevier, 2013). Comparisons were made for articles with 1-10 authors to compare different levels of collaboration. The cut-off point of 10 was chosen as a round number that was large enough to encompass most common collaboration sizes. In some fields, such as high energy physics, very large collaborations are common but these are not of primary interest here. More generally, larger collaborations are more common in some fields than others and so the results will reflect different relative contributions of disciplines for different numbers of authors, which is a limitation of the approach. The first data set was used to compare countries and time, so the geometric MNCS was applied to all 25 subcategories separately in each year. The second data set was used to compare disciplines and countries, so the geometric MNCS was applied to each of the Scopus sections, normalising each subcategory and year separately. Standard normal distribution formulae were used to calculate 95% confidence intervals for the geometric MNCS values before the exponential reverse transformation, since the log transformed data was approximately normally distributed.

The processed data for the results and the R code used to process the raw data can be found online: http://dx.doi.org/10.6084/m9.figshare.1609702.

## 6. Results

The graphs show, as expected, that the average impact of research varies substantially by nation because the country lines tend not to overlap much. A more detailed examination of the graphs is needed to answer the research questions, however.

Figures 1 to 8 confirm that the overall trend is for papers with more authors to have increased average citation impacts. This is clear from the lines for the complete data sets in every year except 2015, for which there are too few citations and so the confidence limits are wide enough that any trends visible are likely to be spurious. The error bars for the complete data set (Figure 8) are narrow enough to be invisible on the graphs for early years and so the positioning of the points can be accepted as being essentially accurate. In this context, any deviations from a smooth line are of interest. By far the largest overall increase is from the average impact of solo authored research to research with two authors in every year before 2014. Additional authors after the second are much less influential in attracting extra citations. Surprisingly, however, the seventh to the tenth authors increase the average citation rate more than do the third to the sixth authors (i.e., the slopes of the All line increases at the seventh author point. This cannot be explained by disciplinary differences (e.g., medical research being more highly cited and tending to have larger teams) because the normalisation process accounts for disciplinary differences in average citation rates. Instead it suggests that in at least some disciplines, and perhaps in most or all, large team sizes tend to produce higher impact research. This could be, for example the result of a substantial number of large-scale medical or epidemiological studies that are particularly highly cited.

There are clear international differences in the citation impact of collaborations. In all data sets before 2015, solo research from Russia has higher average impact than does collaborative research and the overall trend seems to be that Russian national collaboration is somewhat disadvantageous from the perspective of attracting citations, although additional authors after the fourth or fifth start to increase the average citation rate. The robustness of the pattern is compromised by the wide confidence intervals for solo authored research, which contain the data point for two authors in most years. However, the exceptions (2011 and 2012), the similar trend for most years, and the data point for one author being outside of the confidence limits for the data point for two authors in most graphs make this conclusion robust.

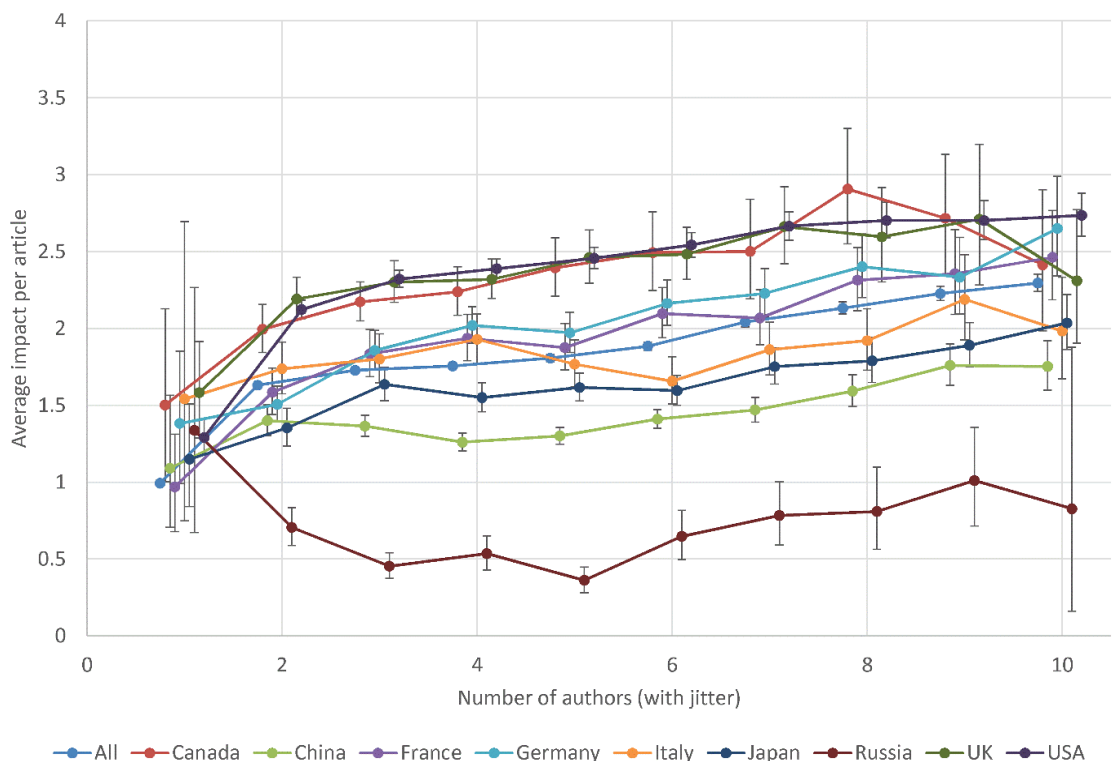

**Figure 1**. GNCI values for articles with all author affiliations from a single country, using journal articles from 25 Scopus categories in 2009. Confidence limits use normal distribution formulae. Jitter has been added to x axis values to prevent the error bars from overlapping (n=170,184).

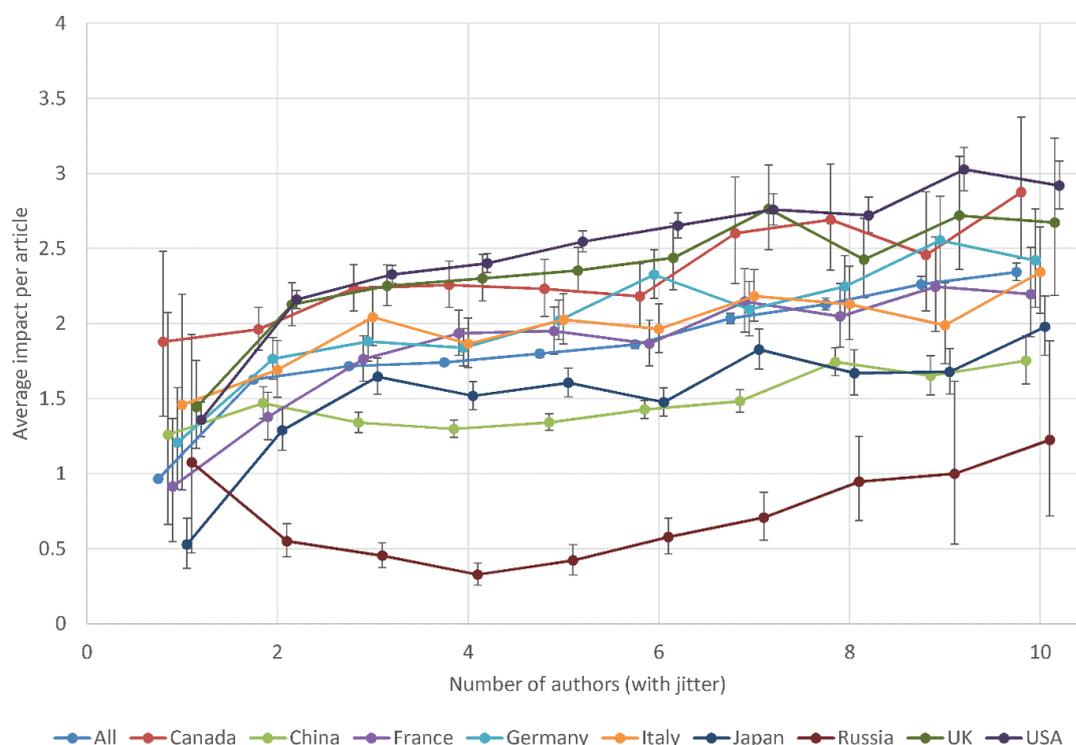

**Figure 2**. GNCI values for articles with all author affiliations from a single country, using journal articles from 25 Scopus categories in 2010. Confidence limits use normal distribution formulae. Jitter has been added to x axis values to prevent the error bars from overlapping (n=174,194).

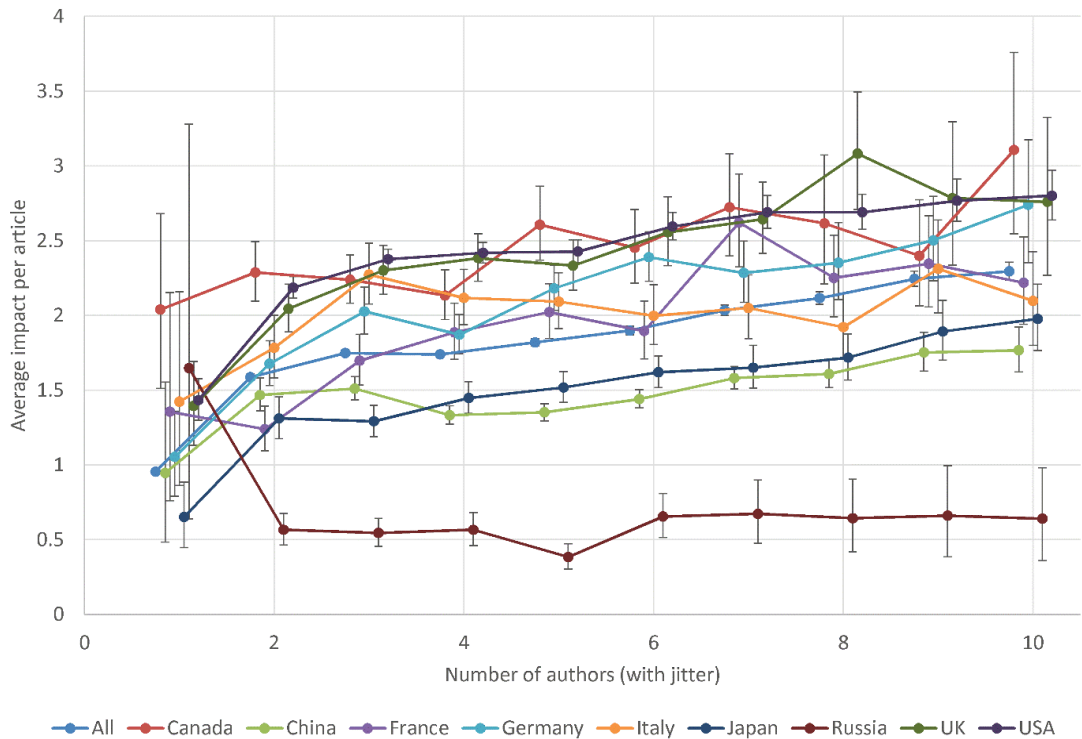

**Figure 3**. GNCI values for articles with all author affiliations from a single country, using journal articles from 25 Scopus categories in 2011. Confidence limits use normal distribution formulae. Jitter has been added to x axis values to prevent the error bars from overlapping (n=181,171).

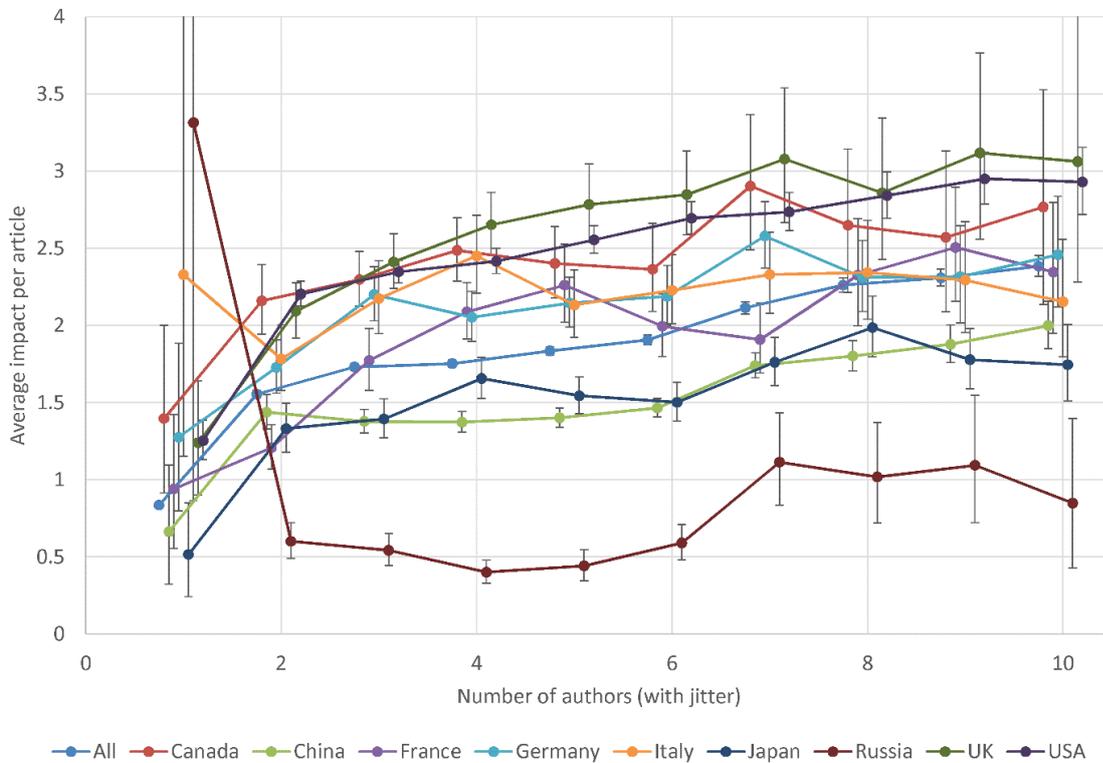

**Figure 4**. GNCI values for articles with all author affiliations from a single country, using journal articles from 25 Scopus categories in 2012. Confidence limits use normal distribution formulae. Jitter has been added to x axis values to prevent the error bars from overlapping (n=187,145).

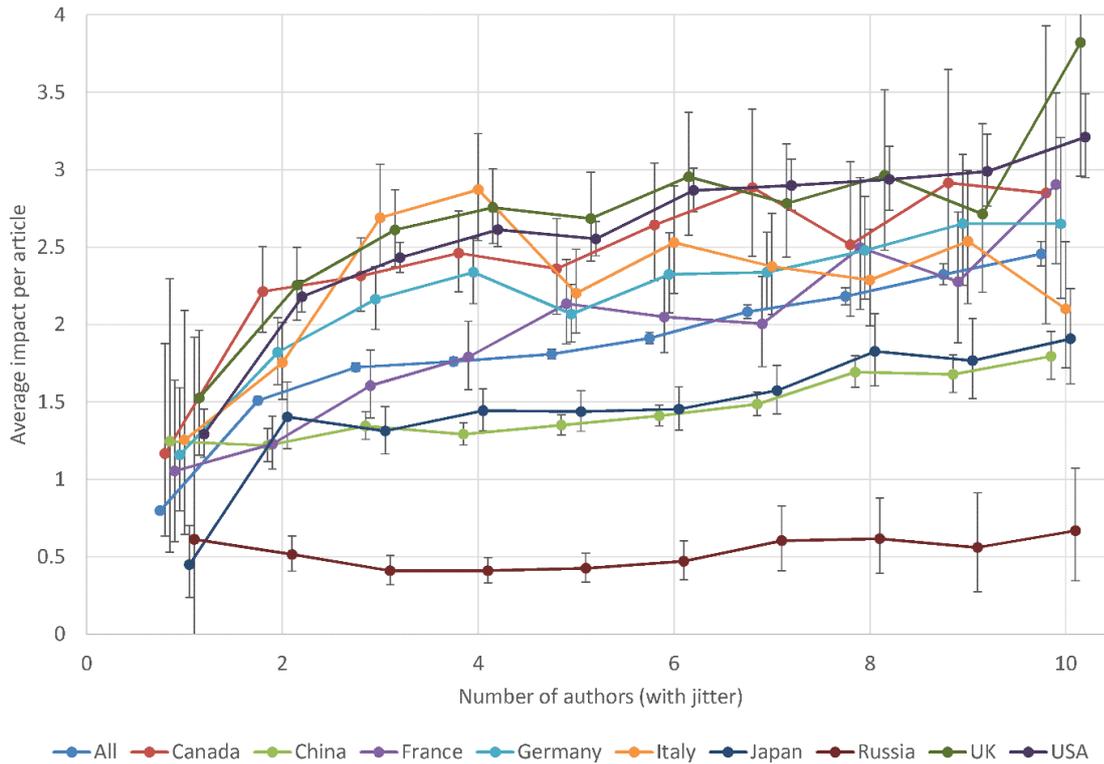

**Figure 5**. GNCI values for articles with all author affiliations from a single country, using journal articles from 25 Scopus categories in 2013. Confidence limits use normal distribution formulae. Jitter has been added to x axis values to prevent the error bars from overlapping (n=189,921).

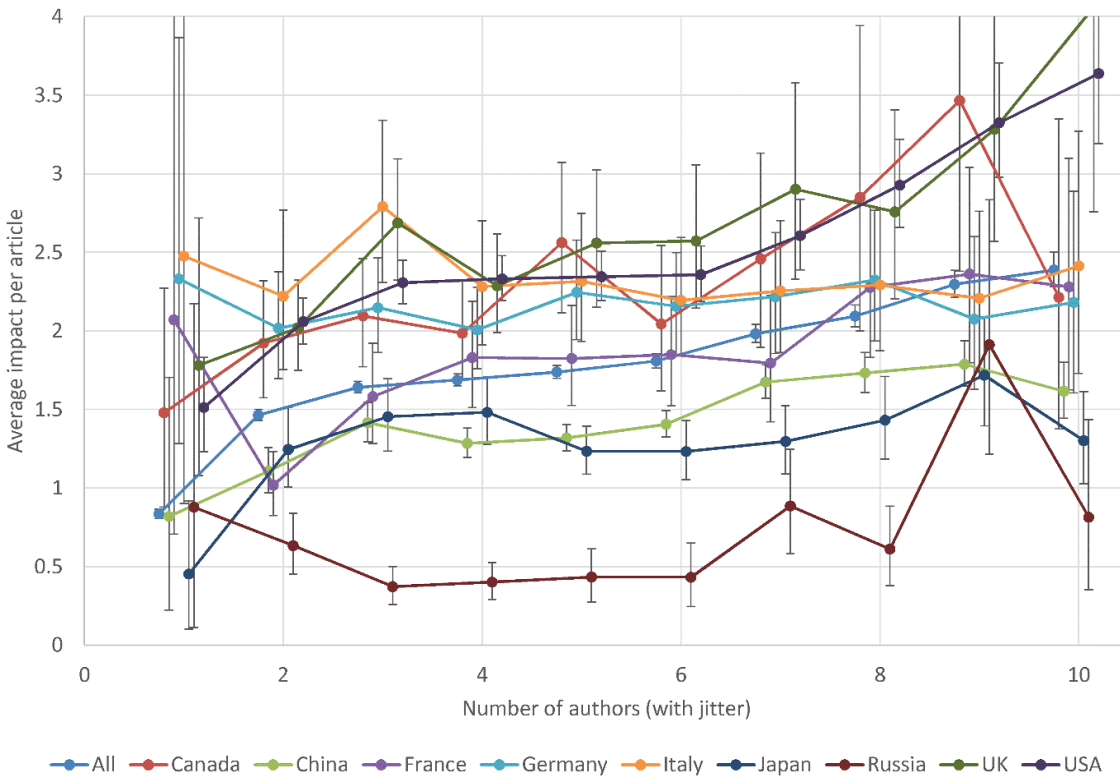

**Figure 6**. GNCI values for articles with all author affiliations from a single country, using journal articles from 25 Scopus categories in 2014. Confidence limits use normal distribution formulae. Jitter has been added to x axis values to prevent the error bars from overlapping (n=177,187).

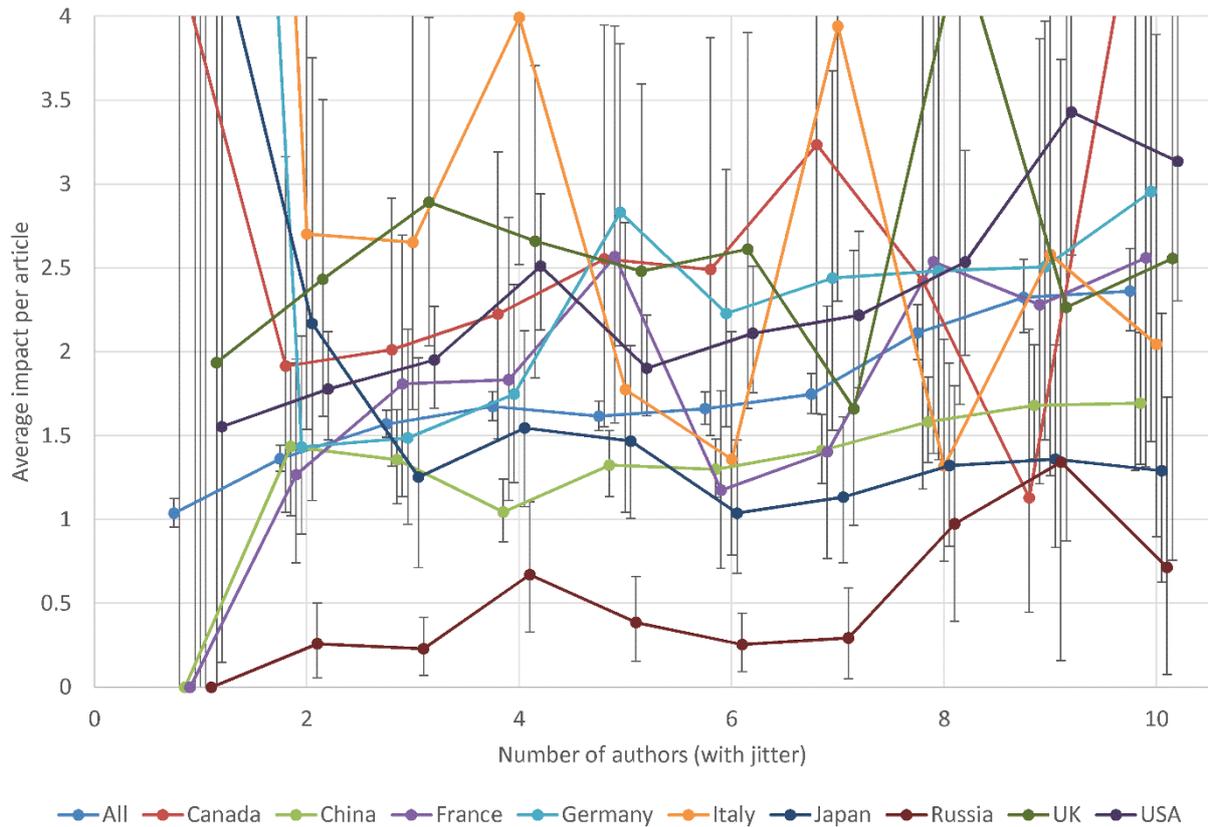

**Figure 7**. GNCI values for articles all with author affiliations from a single country, using journal articles from 25 Scopus categories in 2015. Confidence limits use normal distribution formulae. Jitter has been added to x axis values to prevent the error bars from overlapping (n=149,180).

Changes over time in the average impact of articles with different numbers of co-authors seem to be slight overall, but there are statistically significant changes, as indicated by the confidence intervals from specific years not overlapping with the corresponding points from other years (Figure 8). In particular, 2015 stands out as substantially different from the other years, and 2014 follows it somewhat in being relatively low for 2-7 authors. This seems more likely to be due to the relatively special reasons why articles attract early citations within their publication year than to fundamental changes in the relationship between co-authorship and impact, however. Nevertheless, there are other statistically significant differences between years. For example, the normalised average impact of solo articles 2012-2014 is significantly lower than the average impact of articles 2009-2011, suggesting that there is a trend for solo authorship to attract increasingly few citations, relative to collaborative articles.

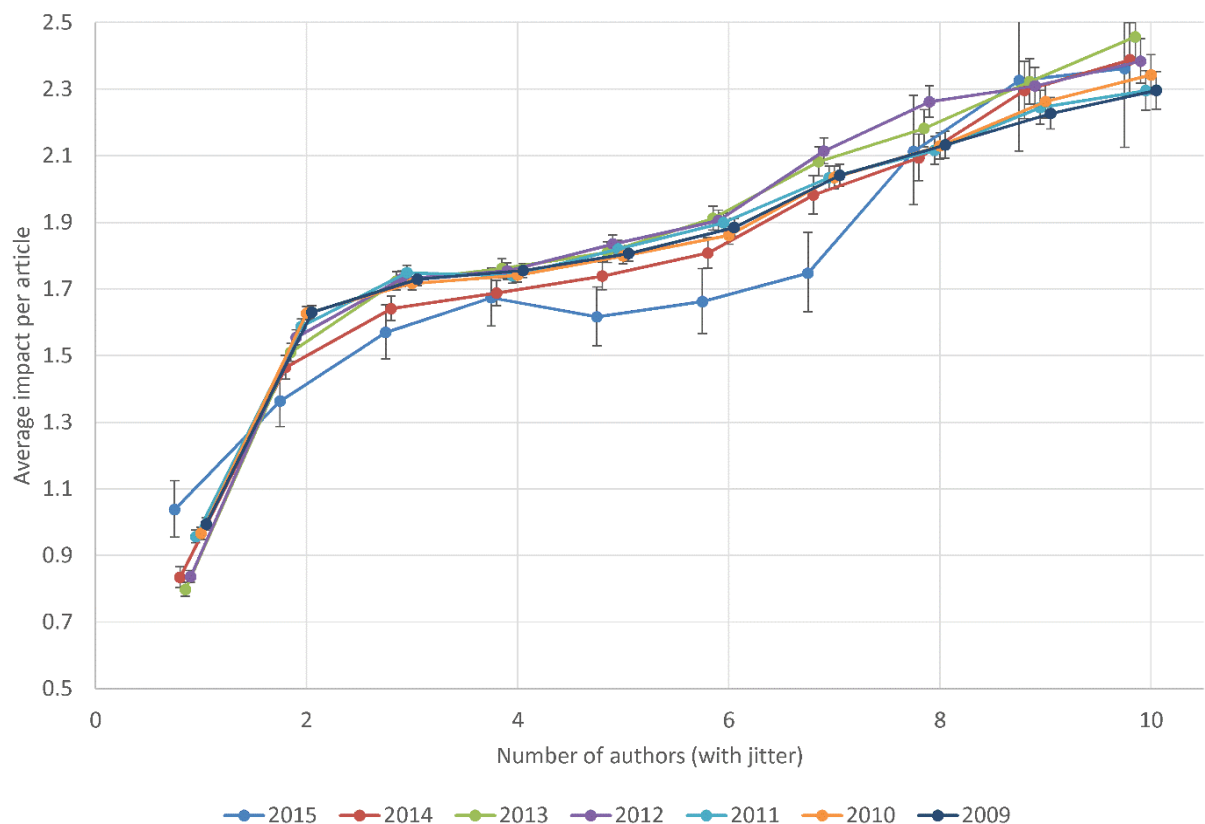

**Figure 8**. GNCI values for journal articles from 25 Scopus categories 2009-2015. Confidence limits use normal distribution formulae. Jitter has been added to x axis values to prevent the error bars from overlapping (n=1,228,982).

In terms of disciplinary differences, there are statistically significant differences in the (field and year normalised) average number of citations for articles with different numbers of authors in the four areas examined (Figure 9). Most strikingly, in the Arts and Humanities, co-authorship has the most substantial association with higher average citation rates. Nevertheless, there are also statistically significant differences in the field normalised average citation rates for most numbers of authors between the other three areas as well. Also of interest is the fact that Pharmacology, Toxicology and Pharmaceutics has a faster increasing rate of citations with increasing co-authorship than does Chemistry, and so large numbers of co-authors are more useful in this area than in Chemistry.

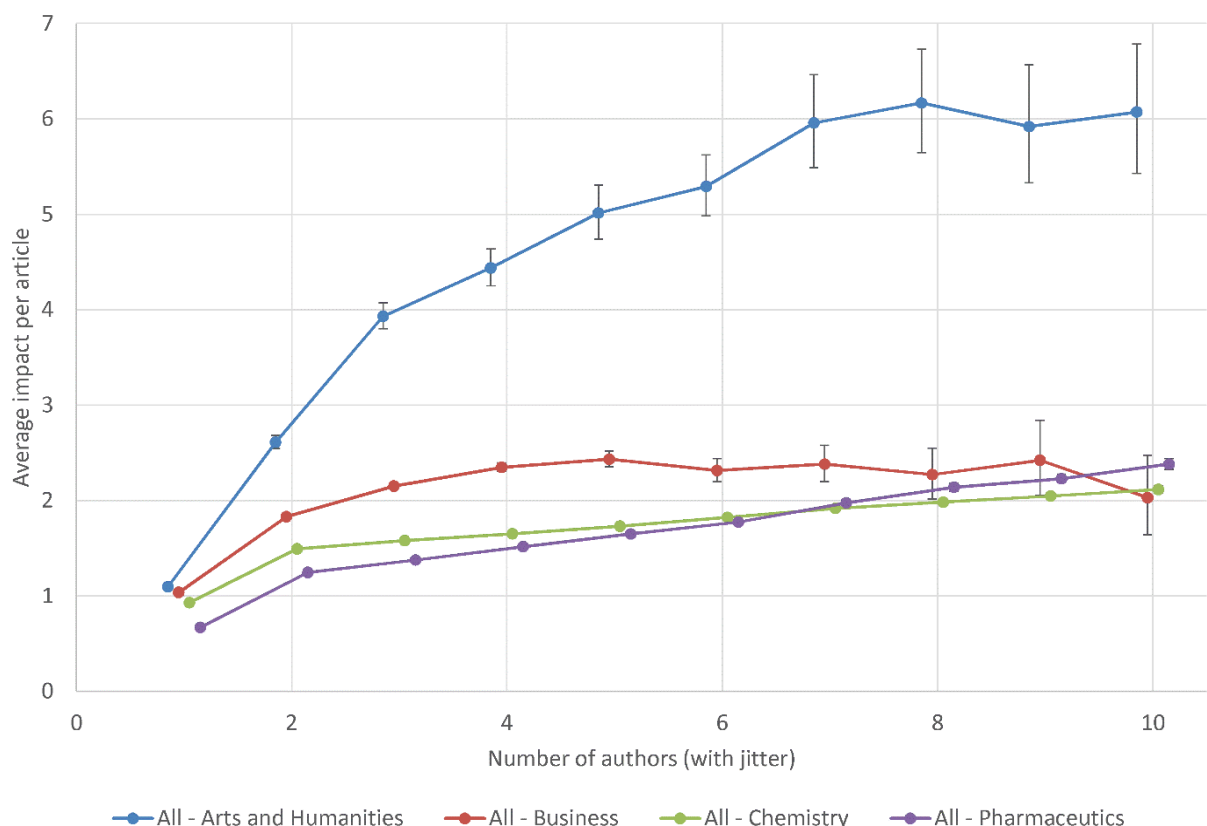

**Figure 9.** GNCI values for journal articles from four broad Scopus categories 2009-2013. Confidence limits use normal distribution formulae. Jitter has been added to x axis values to prevent the error bars from overlapping (n=779,377).

For individual disciplines (Figures 10-13), the country patterns are not always the same as for the multiple discipline graphs (Figures 1-7). Whilst solo authorship associates with higher impact research for Russia overall (Figures 1-5), the reverse is true in Chemistry (Figure 12) and probably also in Pharmacology, Toxicology and Pharmaceutics (Figure 13). This shows that discipline and country are both factors in the association between co-authorship and average citation rates.

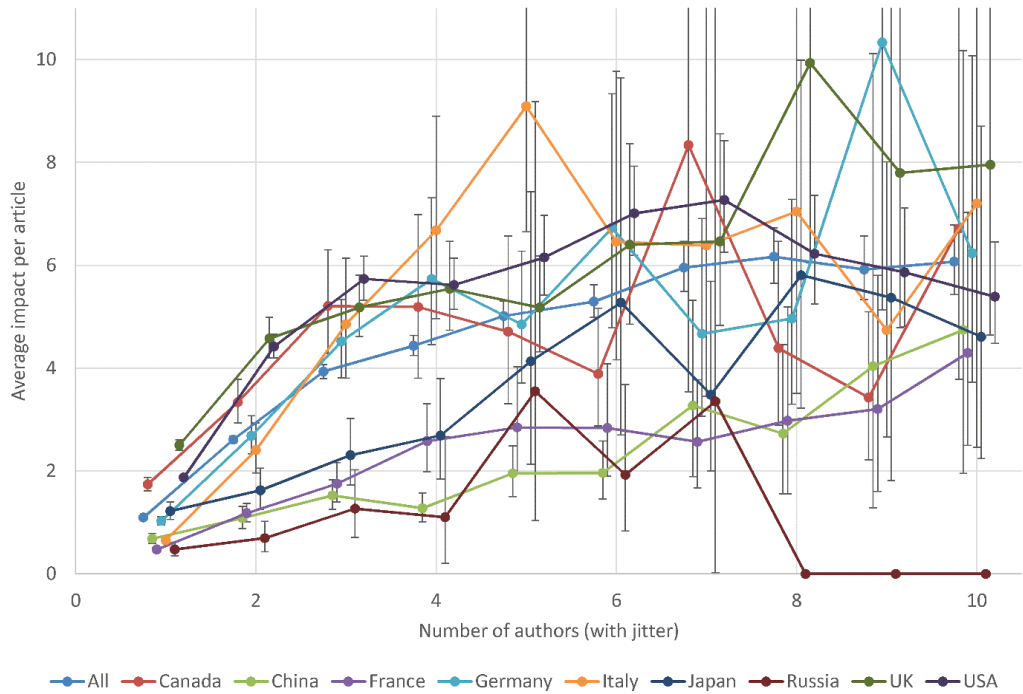

**Figure 10**. GNCI values for articles all author affiliations from a single country, using journal articles from all subcategories of the Scopus broad category of Arts & Humanities, published 2009-2013. Confidence limits use normal distribution formulae. Jitter has been added to x axis values to prevent the error bars from overlapping (n=211,056).

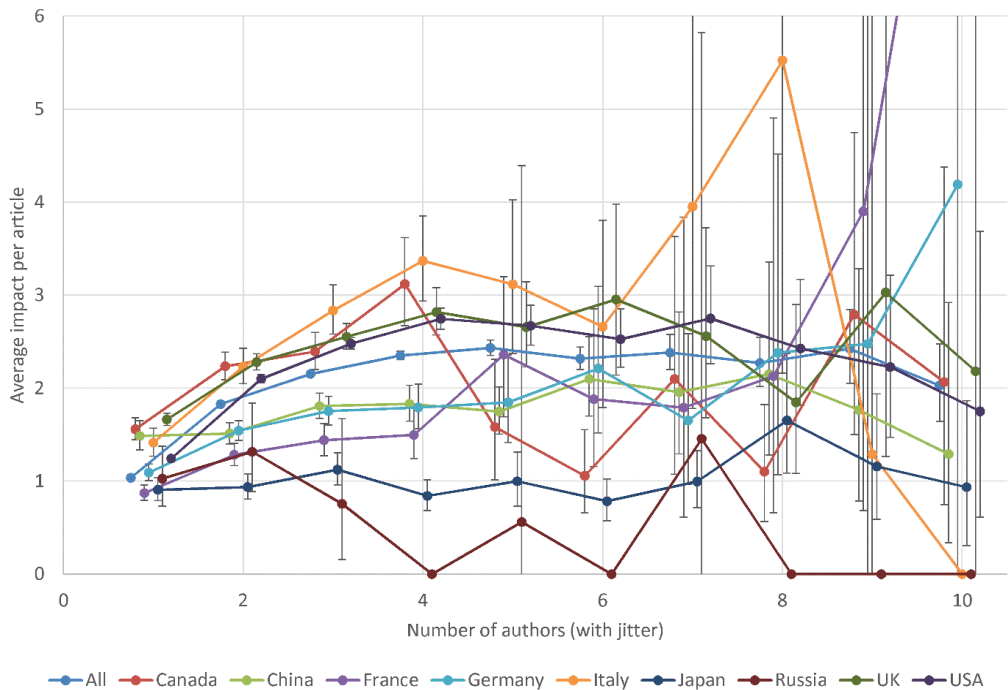

**Figure 11**. GNCI values for articles all author affiliations from a single country, using journal articles from all subcategories of the Scopus broad category of Business, Management and Accounting, published 2009-2013. Confidence limits use normal distribution formulae. Jitter has been added to x axis values to prevent the error bars from overlapping (n=155,346).

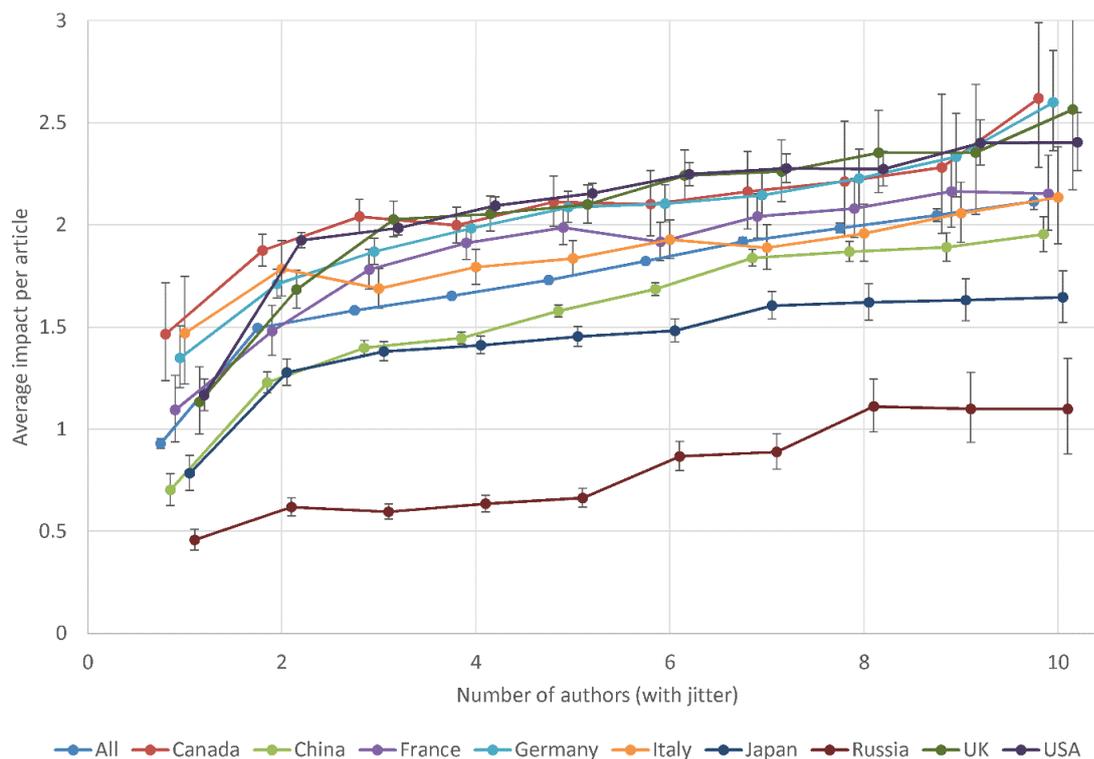

**Figure 12**. GNCI values for articles all author affiliations from a single country, using journal articles from all subcategories of the Scopus broad category of Chemistry, published 2009-2013. Confidence limits use normal distribution formulae. Jitter has been added to x axis values to prevent the error bars from overlapping (n=268,531).

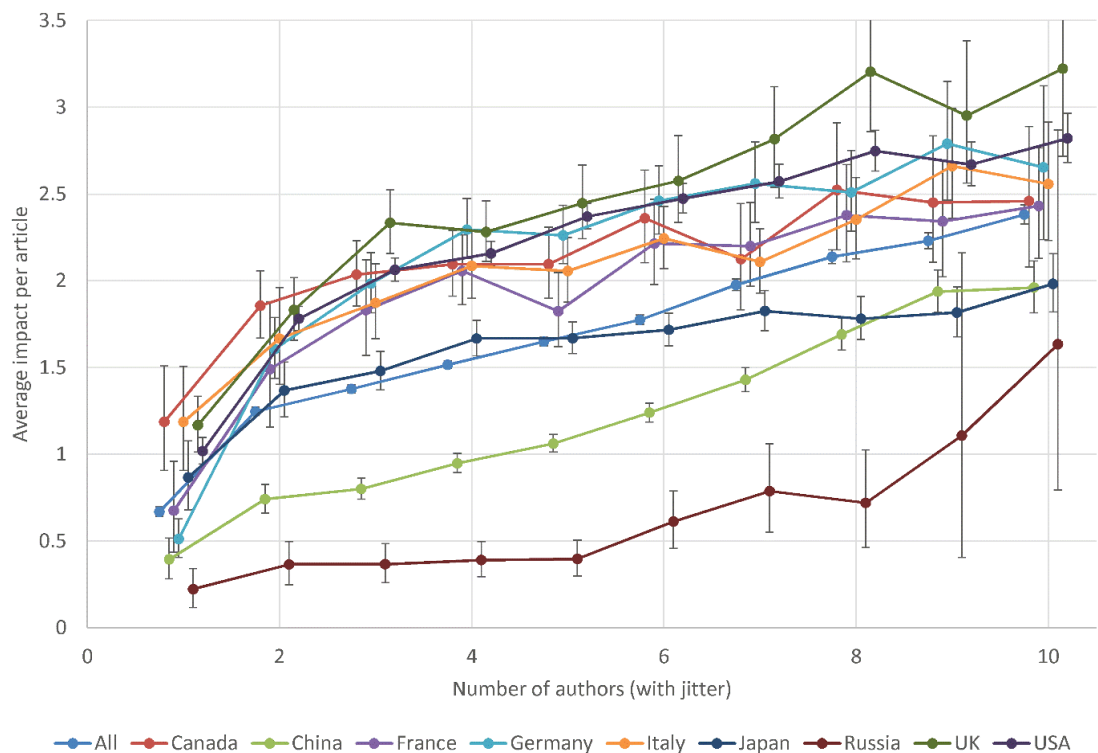

**Figure 13**. GNCI values for articles all author affiliations from a single country, using journal articles from all subcategories of the Scopus broad category of Pharmacology, Toxicology and Pharmaceutics, published 2009-2013. Confidence limits use normal distribution formulae. Jitter has been added to x axis values to prevent the error bars from overlapping (n=144,444).

# 7. Discussion

An important limitation of this study is that the country graphs only included domestic collaborations. Although this is necessary to avoid the possibility that some of the effect of collaboration would derive from the research strength of the collaborating nations, it is an oversimplification because international collaboration is an important aspect of research. As can be seen from additional graphs online, if international collaborations are included then the patterns are broadly the same, however (at the bottom of the "a" versions of worksheets of the supporting files – see the end of the methods section for URL). Whilst the nature of collaborations could have been taken into account with a regression modelling approach (e.g., Thelwall & Maflahi, 2015), this is likely to have generated much wider confidence limits and hence would have given less detailed and less (statistically) robust results.

Another limitation is that the classification system used is that of Scopus, which applies at the level of entire journals rather than individual articles. Since some research is multidisciplinary and any classification system is necessarily a simplification of the complex situation of evolving and overlapping research fields, the results are likely to be affected by the system used (Rafols & Leydesdorff, 2009). In particular, the apparent extreme association between impact and collaboration for the Arts and Humanities may be caused by the collaborative articles being predominantly either multidisciplinary or articles from other areas but published in arts or humanities journals, due to the normal scattering of articles across journals (Bradford, 1934). For example, the journal *Criminology and Public Policy* was categorised within the Literature and Literary Theory subcategory, presumably as a minor classification, and its papers tended to be relatively highly cited and co-authored. Similarly, the *Journal of Archaeological Science* within the History subcategory tended to have highly cited and co-authored papers, presumably being an interdisciplinary research area. This would explain the apparent contradiction that solo work is particularly valued (and particularly common – see Figure 1 in: Larivière, Gingras, & Archambault, 2006) in humanities scholarship, albeit primarily in the form of the monograph (Cronin & La Barre, 2004; Williams, Stevenson, Nicholas, Watkinson, & Rowlands, 2009), for which collaboration associates with fewer citations (Thelwall & Sud, 2014).

There are two further reasons why the trends in the graphs may be misleading to some extent. Since not all subjects collaborate to the same extent, the relative contribution of subjects is different for different numbers of co-authors and so some of the trends could be due to changes in the subject mix rather than to changes in impact due to collaboration. Similarly, some subject areas attract citations at a more rapid rate than others and these subjects would be able to exhibit a greater difference between high and low impact articles for recent years, even after normalisation. They could therefore have a disproportionate influence on the results.

The results should not be taken as proof that collaboration is the *cause* of higher impact research in any case. The data shows an association but not whether there is a cause and effect relationship between the two. There are many possible explanations for the association other than collaboration helping to produce better research. Collaborations may attract more attention to an article if not all of the authors have the same community in which they have name recognition. This may be particularly important for international collaboration (Thelwall & Maflahi, 2015). In addition, researchers that produce higher impact research may naturally attract collaborators or good doctoral students, and so high impact research would cause increased collaboration in subsequent articles. The same directional relationship would also occur if researchers producing high impact research were more likely to be successful in large funding applications that required collaboration (but see: Van den Besselaar & Leydesdorff, 2009). Similarly, more expensive research that requires a larger team may also tend to be more cited (e.g., Luukkonen, Persson, & Sivertsen, 1992), perhaps because it is rarer, more carefully conducted because of the expense, or more important. Finally, it may be that more collaborative research tends to derive from fields that conduct more highly cited research (e.g., see: Franceschet & Costantini, 2010; Wuchty, Jones, & Uzzi, 2007), and so the association between impact and collaboration at the disciplinary level may reflect collaborative types of research tending to be in higher impact research subfields.

The unusual result for Russia, where solo authorship overall associates with higher impact research may be due to different authorship practices within Russia. Perhaps solo research is primarily conducted by senior researchers, whereas research with multiple authors may tend to be the work of junior researchers, with the main co-author being a senior researcher. Evidence from Canada suggests that in many fields the work of PhD students is less cited than average (Larivière, 2011). If this were to be true then co-authorship in Russia may tend to reflect a different type of collaboration than in some other countries. Russia's apparently high level of state support for research related to defence, the disruption caused by the demise of the USSR (Wilson & Markusova, 2004), centralisation of researchers within a few elite research institutes and the importance of international collaboration for highly cited research (Pislyakov & Shukshina, 2014) may also be factors in it having an unusual profile for domestic collaboration.

The unstable results for 2015 (e.g., Figure 7) are due to low citation counts and, in some cases, small numbers of papers. This confirms that international comparisons are not possible for very recent data. This is important to check because policy makers making international comparisons need to have the most recent useful data in order to make an up to date evaluation of policy changes.

## 8. Conclusion

The results confirm that domestic collaboration associates with more highly cited research overall, and that adding a second author to a study makes a much larger difference than each subsequent additional co-author. The results vary by country, however, with solo authorship being advantageous in Russia. Overall, the value of collaboration seems to have increased in the sense that the normalised average impact of solo articles was relatively higher before 2012, although solo articles always have below average impact overall in the data. There are substantial differences between broad disciplines in the extent to which multiple authorship associates with higher impact research as well as in the trend in this relationship (e.g., the overlapping lines in Figure 9). Moreover, the overall trends in impact for domestic collaboration in individual countries vary by discipline, for example, with Russian domestic collaboration associating with higher impact research in Chemistry but not overall.

From the perspective of the geometric MNCS, its more precise estimation of the average citation impact of collections of articles from multiple fields and/or years has allowed a fine grained examination of patterns in the results, with conclusions that are probably more robust than could have been made with previous indicators. In addition, the confidence limits associated with the indicator have allowed the level of confidence in the conclusions to be estimated. Hence it seems to be a useful and practical new indicator for comparing the citation impact of sets of articles from multiple fields and/or years.